%% file: main.tex
\documentclass[letterpaper, 10 pt, conference]{ieeeconf}
\usepackage[utf8]{inputenc}

\IEEEoverridecommandlockouts

\usepackage{balance}

\usepackage{multirow}

\usepackage{amsmath,mathtools}  %
\usepackage{amsfonts,amssymb}
\usepackage{bbm}

\usepackage{xspace}    %

\usepackage{graphicx}
\graphicspath{{figs/}}

\usepackage{subcaption}
\usepackage{caption}

\usepackage{hyperref}
\hypersetup{colorlinks=true, linkcolor= blue, allcolors=blue, citecolor = blue, filecolor=black, urlcolor=blue}

\usepackage{todonotes}
\usepackage{comment}

\usepackage{tikz}
\usetikzlibrary{arrows,positioning}   %

\usepackage{pgfplots}\pgfplotsset{compat=1.9}
\usepackage{grffile}
\pgfplotsset{compat=newest}  %
\usetikzlibrary{plotmarks}
\usetikzlibrary{arrows.meta}
\usepgfplotslibrary{patchplots}
\usepgfplotslibrary{groupplots}

\pgfplotsset{compat=1.15}
\usepackage{mathrsfs}
\usetikzlibrary{arrows}

\usepackage{tabularx}

\usepackage[ruled]{algorithm}
\usepackage[noend]{algpseudocode}

\usepackage{diagbox}
\newcolumntype{K}[1]{>{\centering\arraybackslash}p{#1}}

\makeatletter
\newcommand{\multiline}[1]{%
  \begin{tabularx}{\dimexpr\linewidth-\ALG@thistlm}[t]{@{}X@{}}
    #1
  \end{tabularx}
}
\makeatother

\usepackage[per-mode=symbol]{siunitx}

\usepackage{cite}  %
\usepackage{array}  %
\usepackage{booktabs}           %

\newcounter{remark}
\newenvironment{remark}{%
\par\vspace{3pt}\noindent\refstepcounter{remark}\textbf{Remark~\theremark:}}%
{\par\endtrivlist\unskip}

\newtheorem{definition}{Definition}

\newcounter{problem}
{\par\endtrivlist\unskip}

\usepackage[extramath,probability,reviewmark]{mylatexdefs}

\input{defs}

\usepackage{stfloats}%

\begin{document}

\title{\LARGE \bf Combining Graph Attention Networks and Distributed Optimization for Multi-Robot Mixed-Integer Convex Programming}
\author{Viet-Anh Le, {\IEEEmembership{Student Member, IEEE}}, 
Panagiotis Kounatidis, {\IEEEmembership{Student Member, IEEE}}, \\
and Andreas A. Malikopoulos, {\IEEEmembership{Senior Member, IEEE}}
\thanks{This research was supported in part by NSF under Grants CNS-2149520, CMMI-2348381, IIS-2415478, and in part by Mathworks.}
\thanks{Viet-Anh Le is with the Department of Mechanical Engineering, University of Delaware, Newark, DE 19716 USA, and also with the Systems Engineering Field, Cornell University, Ithaca, NY 14850 USA (email: {\tt\small vl299@cornell.edu}).}
\thanks{Panagiotis Kounatidis is with the Systems Engineering Field, Cornell University, Ithaca, NY 14850 USA (email: {\tt\small pk586@cornell.edu}).}
\thanks{Andreas A. Malikopoulos is with School of Civil and Environmental Engineering, Cornell University, Ithaca, NY 14853 USA, and also with the Systems Engineering Field, Cornell University, Ithaca, NY 14850 USA (email: {\tt\small amaliko@cornell.edu}).}
}

\maketitle

\begin{abstract}
In this paper, we develop a fast mixed-integer convex programming (MICP) framework for multi-robot navigation by combining graph attention networks and distributed optimization. We formulate a mixed-integer optimization problem for receding horizon motion planning of a multi-robot system, taking into account the surrounding obstacles. To address the resulting multi-agent MICP problem in real time, we propose a framework that utilizes heterogeneous graph attention networks to learn the latent mapping from problem parameters to optimal binary solutions. Furthermore, we apply a distributed proximal alternating direction method of multipliers algorithm for solving the convex continuous optimization problem. We demonstrate the effectiveness of our proposed framework through experiments conducted on a robotic testbed.
\end{abstract}

\section{Introduction}

Mixed-integer optimization-based control plays a crucial role in various real-world applications that involve discrete decision-making, including autonomous driving \cite{quirynen2024real}, traffic signal coordination with connected automated vehicles \cite{le2024distributed}, vehicle routing \cite{bang2021AEMoD}, multi-robot pick-up and delivery \cite{camisa2022multi} or navigation in crowds \cite{le2023multirobot}, and motion planning and goal assignment for robot fleets \cite{salvado2018motion} to name a few.
As a result, there has been growing interest in developing fast algorithmic frameworks to efficiently solve or approximate mixed-integer programs in real time.
For receding horizon motion planning and control where the sampling period is limited, heuristic techniques have been developed to find approximate solutions, in which optimality may be sacrificed for real-time practicality.
Some common heuristic techniques include rounding schemes \cite{bestehorn2019switching}, the feasibility pump \cite{zhang2021feasibility}, and approximate dynamic programming \cite{stellato2016high}.

Recently, there has been significant interest in leveraging machine learning to accelerate algorithms for mixed-integer programs, with a particular focus on mixed-integer convex programs (MICPs).
MICPs refer to a class of problems that become convex once the integer feasibility (integrality) constraints are relaxed, including mixed-integer linear programs (MILPs) and mixed-integer quadratic programs (MIQPs).
In receding horizon motion planning and control, MICPs at each time step are generally formulated as parametric MICPs, where certain problem parameters, such as initial conditions, vary over time.
Supervised learning can then approximate the latent mapping from problem parameters to binary solutions, transforming non-convex MICPs into convex programs that can be solved efficiently online.
For example, classification models based on feedforward neural networks (NNs) \cite{cauligi2020learning,cauligi2021coco,bertsimas2022online}, imitation learning \cite{srinivasan2021fast}, and long short-term memory networks \cite{cauligi2022prism} have been used to learn optimal binary solutions offline.
Bertsimas \etal \cite{bertsimas2023prescriptive} proposed a prescriptive algorithm instead of classification algorithms to take into account all available decision options.
Russo \etal \cite{russo2023learning} trained an NN to predict an integer solution along with a set of linear programs whose feasible sets partition the MILP's feasible set.
For mixed-integer nonlinear programming, Tang \etal \cite{tang2024learning} proposed a framework with integer correction layers to ensure integrality and a projection step to improve solution feasibility.

Prior work to date has focused on single-agent systems.
However, many real-world applications where MICP can be utilized involve multi-agent systems, \eg \cite{le2024distributed, camisa2022multi, salvado2018motion}.
The learning frameworks for single-agent systems do not readily extend to multi-agent systems due to several reasons.
First, the optimization problem formulation depends on hyper-features of the multi-agent system, such as the number of agents and the pairs of agents sharing coupling constraints, which may be time-varying.
Second, the optimal solution for each agent is influenced not only by its features but also by the features and solutions of other agents, introducing complex interdependencies.
Considering multi-robot navigation as an example, where binary variables can be employed in collision avoidance constraints, the optimal trajectory for each robot—including the binary decisions for collision avoidance—depends not only on its current state and the obstacle's parameters but also on the trajectories of other robots.
Therefore, in this paper, we aim to develop a framework for learning the optimal binary solutions in multi-robot navigation based on \emph{heterogeneous graph attention (GAT)} networks.
We first model the multi-robot system with obstacles as an \emph{heterogeneous graph}, where nodes represent both robots and obstacles, while edges indicate which pairs of agents share collision avoidance coupling constraints.
Then, we design an NN consisting of an encoder that leverages a heterogeneous GAT network to incorporate graph-structured data and capture the latent dependencies between nodes and a decoder that predicts the optimal binary strategy for each edge in the graph.
Furthermore, to reduce the solving time for the multi-agent convex program when the number of agents is large, we adopt a distributed proximal ADMM \cite{deng2017parallel}, which enables \emph{parallel computation} in a multi-agent optimization either with classical or nonclassical information structures \cite{Malikopoulos2021}.
Thus, the proposed framework takes advantage of both offline supervised learning and online distributed optimization for solving multi-agent MICPs in real time.

The rest of this paper is organized as follows.
Section~\ref{sec:prb} formulates a multi-agent MICP for multi-robot navigation given surrounding obstacles.
Sections~\ref{sec:gnn} and \ref{sec:opt} present an NN architecture based on heterogeneous GAT networks to learn the optimal binaries and the proximal ADMM for solving the convex program in a distributed manner, respectively.
Section~\ref{sec:exp} validates the proposed framework with experimental results, and Section~\ref{sec:conc} provides some concluding remarks.

\section{Problem Formulation}
\label{sec:prb}
In this section, we first formulate a MICP for multi-robot navigation in the presence of stationary obstacles.
Then, we recast the problem as a multi-agent parametric MICP, allowing us to apply a supervised learning method to accelerate solving the problem.

\subsection{Multi-Robot Navigation}

\begin{figure}[tb]
\centering
\vspace{7pt}
\includegraphics[width=0.48\textwidth]{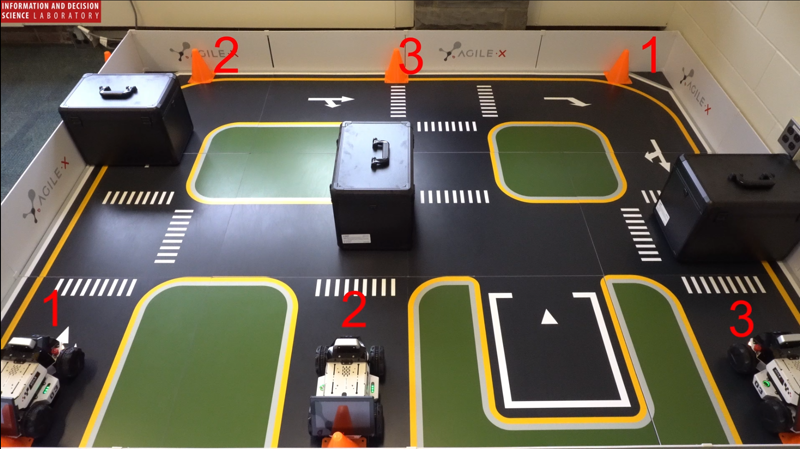}
\caption{Multi-robot navigation problem with stationary obstacles.}
\label{fig:env}
\vspace{-10pt}
\end{figure}

We consider the multi-robot navigation problem with stationary obstacles, as illustrated in Fig.~\ref{fig:env}.
We assume that the obstacles are stationary and represented by rectangles. 
However, the formulation can be readily extended to any polyhedral obstacles.
In this problem, the robots are required to navigate from their initial locations to assigned goals, while avoiding collisions with both the obstacles and other robots.
Let $N_R \in \ZZplus$ and $N_O \in \ZZplus$  be the numbers of robots and obstacles, respectively.
Next, we introduce the definitions of the entities in our problem setup.

\begin{definition}
(Robots)
Let $\RRR = \{1, \dots, N_R\}$ be the set of robots and $\EEE_R \subset \RRR \times \RRR$ be the edge set for pairs of robots.
We assume that $\EEE_R$ is bidirectional.
A pair of robots form an edge if they share collision avoidance constraints, which will be defined later.
For each \robot{i}, $i \in \RRR$, let $\NNN_i = \{j \in \VVV \;|\; (i,j) \in \EEE_R\}$ be the set of its neighboring robots.
\end{definition}

\begin{definition}
(Obstacles)
Let $\OOO = \{1, \dots, N_O\}$ be the set of obstacles and $\EEE_{RO} \subset \RRR \times \OOO$ be the edge set for pairs of robot-obstacle.
A robot-obstacle pair forms an edge if the robot needs to avoid collision with the obstacle.
\end{definition}

Let $t \in \ZZplus$ be the current time step.
Next, we formulate a MICP over a finite control horizon of length $H \in \ZZplus$ starting from time $t$.
At every time step $k \in \ZZplus$, let $\bb{p}_{i}(k) = [p^x_{i}(k),p^y_{i}(k)]^\top \in \RR^2$, $\bb{v}_{i}(k) = [v^x_{i}(k),v^y_{i}(k)]^\top \in \RR^2$, and $\bb{u}_{i}(k) = [u^x_{i}(k),u^y_{i}(k)]^\top \in \RR^2$ be the vectors of positions, velocities, and accelerations for each \robot{i}, respectively.
Let $\bb{x}_{i}(k) = [\bb{p}_{i}(k),\bb{v}_{i}(k)]^\top$ be the state vector of \robot{i}.  
The dynamics of each robot are governed by a discrete-time double-integrator model as follows
\begin{equation}
\label{eq:dynamics}
\begin{split}
\bb{p}_{i} (k+1) &= \bb{p}_{i} (k) + \tau \bb{v}_{i} (k) + \frac{1}{2} \tau^2 \bb{u}_{i} (k), \\
\bb{v}_{i} (k+1) &= \bb{v}_{i} (k) + \tau \bb{u}_{i} (k),
\end{split}
\end{equation}
where $\tau \in \RRplus$ is the sampling time period. 
We assume that the states and control inputs of robots $i\in\RRR$ are subjected to the following bound constraints: %
\begin{equation}
\label{eq:bound}
\begin{split}
p^x_{\min} & \le p^x_{i} (k) \le p^x_{\max},\\    
p^y_{\min} & \le p^y_{i} (k) \le p^y_{\max},\\    
-v_{\max} & \le v^x_{i} (k), v^y_{i} (k) \le v_{\max},\\    
-a_{\max} & \le u^x_{i} (k), u^y_{i} (k) \le a_{\max},
\end{split}
\end{equation}
where $[p^x_{\min}, p^x_{\max}, p^y_{\min}, p^y_{\max}]^\top \in \RR^4$ is the boundary of the environment, $v_{\max} \in \RRplus$ and $a_{\max} \in \RRplus$ are the maximum speed and acceleration of the robots, respectively.
We compactly express \eqref{eq:bound} as $\bbsym{x}_i(k) \in \XXX$ and $\bbsym{u}_i(k) \in \UUU$.

We employ the mixed-integer formulation to formulate robot-robot and robot-obstacle collision avoidance constraints.
First, collision avoidance between a pair of robots can be achieved by the following constraint:
\begin{equation}
| p^x_{i} (k) - p^x_{j} (k) | \ge 2 d_{\min} \;
\mathrm{OR} \; %
| p^y_{i} (k) - p^y_{j} (k) | \ge 2 d_{\min},
\end{equation}
where $d_{\min}$  is the radius of the safe circle for each robot.
This collision avoidance constraint can be represented equivalently using big-$M$ formulation as follows \cite{alrifaee2014centralized}
\begin{equation}
\label{eq:robot_ca}
\begin{split}
p^x_i(k) - p^x_j(k) & \ge 2 d_{\min} - M b_{1,ij}(k), \\
p^x_j(k) - p^x_i(k) & \ge 2 d_{\min} - M b_{2,ij}(k), \\
p^y_i(k) - p^y_j(k) & \ge 2 d_{\min} - M b_{3,ij}(k), \\
p^y_j(k) - p^y_i(k) & \ge 2 d_{\min} - M b_{4,ij}(k),
\end{split}
\end{equation}
where $b_{1,ij}(k)$, $b_{2,ij}(k)$, $b_{3,ij}(k)$ and $b_{4,ij}(k)$ are binary variables satisfing $b_{1,ij}(k) + b_{2,ij}(k) + b_{3,ij}(k) + b_{4,ij}(k) \le 3$ and $M$ is a sufficiently large positive constant.
Note that we always have the relation $b_{1,ij}(k) = b_{2,ji}(k)$, $b_{2,ij}(k) = b_{1,ji}(k)$, $b_{3,ij}(k) = b_{4,ji}(k)$, and $b_{4,ij}(k) = b_{3,ji}(k)$.
We consider inter-robot collision avoidance constraints only for robots whose current distance is below a given threshold $d_\mathrm{prox} > 0$ to prevent unnecessary binary variables and constraints, \ie
\begin{equation}
(i,j) \in \EEE_R \, \iff \, \norm{\bb{p}_i(t) - \bb{p}_j(t)} \le d_\mathrm{prox}.
\end{equation}
Similarly, we can formulate mixed-integer constraints for collision avoidance with obstacles as follows
\begin{equation}
\label{eq:obs_ca}
\begin{split}
& \begin{multlined}
\cos \alpha_o (p^x_i(k) - p^x_o) + \sin \alpha_o (p^y_i(k) - p^y_o) \\ \ge L_o + d_{\min} - M b_{1,io}(k),
\end{multlined} \\
& \begin{multlined}
- \sin \alpha_o (p^x_i(k) - p^x_o) + \cos \alpha_o (p^y_i(k) - p^y_o) \\ \ge W_o +  d_{\min} - M b_{2,io}(k), \end{multlined} \\
& \begin{multlined}
-\cos \alpha_o (p^x_i(k) - p^x_o) - \sin \alpha_o (p^y_i(k) - p^y_o) \\ \ge L_o + d_{\min} - M b_{3,io}(k), \end{multlined} \\
& \begin{multlined} \sin \alpha_o (p^x_i(k) - p^x_o) - \cos \alpha_o (p^y_i(k) - p^y_o) \\ \ge W_o + d_{\min} - M b_{4,io}(k), \end{multlined} \\
\end{split}
\end{equation}
where $b_{1,io}(k)$, $b_{2,io}(k)$, $b_{3,io}(k)$ and $b_{4,io}(k)$ are binary variables satisfying $b_{1,io}(k) + b_{2,io}(k) + b_{3,io}(k) + b_{4,io}(k) \le 3$, 
$[p^x_o, p^y_o]^\top$ is center location, $\alpha_o$ is the rotation angle, and $2L_o$ and $2W_o$ are the length and width of \obstacle{o}$\in \OOO$.

The individual objective for each robot is to reach the goal, \ie minimize the distance to the goal, while maintaining the minimum effort.
Thus, it can be given by a weighted sum of multiple terms as follows
\begin{equation}
\begin{multlined}
\underset{\substack{\bb{x}_i(k+1) \in \XXX,\\ \bb{u}_i (k) \in \UUU}}{\minimize} \;\; \bar{c}_i \big(\bb{x}_{i} (t+H)\big) + \!\sum_{k=t}^{t+H-1} c_i \big(\bb{u}_{i}(k), \bb{x}_{i}(k) \big),
\end{multlined}
\end{equation}
where
\begin{equation}
\begin{split}
&\bar{c}_i (\bb{x}_{i}(t+H)) = \omega_{pt} \big( \bb{p}_i(t+H) - \bb{p}_i^{\rm{g}} \big)^\top \big( \bb{p}_i(t+H) - \bb{p}_i^{\rm{g}} \big), \\
&c_i (\bb{u}_{i}(k), \bb{x}_{i}(k)) = \omega_p \big( \bb{p}_i(k) - \bb{p}_i^{\rm{g}} \big)^\top \big( \bb{p}_i(k) - \bb{p}_i^{\rm{g}} \big) \\
& \hspace{25mm} + \omega_u \bb{u}_i(k)^T \bb{u}_i(k),    
\end{split}
\end{equation}
with $\bb{p}_i^{\rm{g}}$ being the vector of goal positions for \robot{i}, while $\omega_{pt}$, $\omega_p$, and $\omega_u$ are positive weights.

Therefore, the MICP at each time step $t$ is presented as follows
\begin{align}
\label{eq:mpc}
\begin{split}
&\underset{\substack{\bb{x}_i(k+1) \in \XXX,\\ \bb{u}_i (k) \in \UUU}}{\minimize} \;\; \bar{c}_i (\bb{x}_{i}(t+H)) + \! \sum_{k=t}^{t+H-1} c_i (\bb{u}_{i}(k), \bb{x}_{i}(k)),
\\
& \subjectto 
\\
& \quad \eqref{eq:dynamics},\; \forall\, k = t, \dots, t+H-1,
\\
& \quad \eqref{eq:robot_ca},\; \forall\, (i,j) \in \EEE_R,\; \forall\, k = t+1, \dots, t+H,
\\
& \quad \eqref{eq:obs_ca},\; \forall\, (i,o) \in \EEE_{RO},\; \forall\, k = t+1, \dots, t+H,
\\
& \text{given:} 
\\
& \quad \bb{p}_i(t), \; \bb{v}_i(t), \; \forall\, i \in \RRR. \\
\end{split}
\end{align}

\subsection{ Parametrized MICP Formulation}

Next, we express the mixed-integer optimization problem for multi-robot navigation, presented in the previous section, in a general parametric MICP form.
Let $\bb{\theta}_i = [\bb{p}_{i}(t)^\top, \bb{p}_{i}^{\mathrm{g}\top}]^\top$ and $\bb{\theta}_o = [p^x_{o}, p^y_{o}, \alpha_o, L_o, W_o]^\top$ be the vectors of parameters of \robot{i} and \obstacle{o}.
As a result, the objective function and constraints in \eqref{eq:mpc} can be represented as functions of the parameter vectors. 
For ease of notation, we denote $\bb{x}_{i,k} = \bb{x}_{i} (t+k)$, $\bb{u}_{i,k} = \bb{u}_{i} (t+k)$, $\bb{\delta}_{io, k} = [b_{m,io}(t+k)]_{m=1,\dots,4}^\top$, and $\bb{\delta}_{ij, k} = [b_{m,ij}(t+k)]_{m=1,\dots,4}^\top$ where the current time $t$ is dropped.
We denote $\bb{x}_i$, $\bb{u}_i$, $\bb{\delta}_{io}$, and $\bb{\delta}_{ij}$ as the concatenated vectors collecting the corresponding variables over the control horizon.
The parametric MICP formulation is given as follows
\begin{subequations}
\label{eq:paraMPC}
\begin{align}
&
\begin{multlined}
\underset{\substack{\bb{x}_i \in \XXX^{H+1},\\ \bb{u}_i \in \UUU^H}}{\minimize} \;\; \sum_{i \in \RRR} \Big( \bar{c}_i (\bb{x}_{i,H}; \bb{\theta}_i) + \sum_{k=0}^{H-1} c_i (\bb{u}_{i,k}, \bb{x}_{i,k}; \bb{\theta}_i) \Big),
\end{multlined}
\label{eq:MPC:obj}\\
& \text{subject to: } \nonumber \\
& \quad \bb{x}_{i,0} = \bb{x}_{i,\rm{init}} (\bb{\theta}_i), \label{eq:paraMPC-init} \\
& \quad \bb{x}_{i, k+1} = \bb{f}_i (\bb{x}_{i, k}, \bb{u}_{i, k}), \label{eq:paraMPC-dyn} \\
& \quad \bb{g}_{io} (\bb{x}_{i, k}, \bb{\delta}_{io,k}; \bb{\theta}_o) \le 0, \label{eq:paraMPC-obs} \\
& \quad \bb{h}_{ij} (\bb{x}_{i, k}, \bb{x}_{j, k}, \bb{\delta}_{ij, k}) \le 0, \label{eq:paraMPC-rob} \\
& \quad \bb{\delta}_{io,k} \in \{0,1\}^{4},\; \bb{\delta}_{ij,k} \in \{0,1\}^{4}.
\end{align}
\end{subequations}
In \eqref{eq:paraMPC}, constraints \eqref{eq:paraMPC-init}, \eqref{eq:paraMPC-dyn}, \eqref{eq:paraMPC-obs}, and \eqref{eq:paraMPC-rob} represent the initial conditions, system dynamics, obstacle collision avoidance, and inter-robot collision avoidance, respectively.
The state $\bb{x}_{i}$ and control $\bb{u}_{i}$ are the continuous decision variables, while $\bb{\delta}_{io}$ and $\bb{\delta}_{ij}$ are the binary decision variables.
It can be observed that if the binary decision variables $\bb{\delta}_{io}$ and $\bb{\delta}_{ij}$ are fixed, then the remaining problem is convex and can be solved efficiently by any distributed optimization algorithm for convex optimization.
Moreover, the optimal solution at each time step depends on the problem parameters $\bb{\theta}_i$ and $\bb{\theta}_o$.
Thus, an interesting approach to solving problems involves learning a map between the vector of problem parameters and the optimal binaries by supervised learning.
Specifically, we learn and predict the optimal binaries $\{ \bb{\theta}_{io}^* \}_{(i,o) \in \EEE_{RO}}$ and $\{ \bb{\delta}_{ij}^* \}_{(i,j) \in \EEE_R}$.

\textbf{Challenges and our approach:} There are several challenges in learning the optimal binary variables in our multi-robot navigation problem.
First, we need to predict binary variables that describe various relations, such as collision avoidance among robots or between robots and obstacles.
Second, the optimal binaries of each edge depend not only on the features of the two connecting nodes but also on the binaries of some other nodes, such as nearby robots.
Finally, the number of binary variables used changes over time and varies across different scenarios.
In other words, the binary variables in our problem are structured based on graph hyper-features, including the number of nodes and pairs of agents sharing coupling constraints.
Therefore, it is necessary to incorporate the graph data as input into the learning model.
In the next section, we develop a framework that addresses the above issues by leveraging a heterogeneous graph attention network.

\section{Learning with Heterogeneous Graph Attention Networks}
\label{sec:gnn}

The main idea is to use graph attention neural networks to aid in solving multi-agent MICPs.
The framework consists of both offline training and an online optimization algorithm.
In particular, we design a graph attention network to learn the best binary strategy from data offline, and then solve convex continuous optimization problems online using a distributed optimization algorithm.
To this end, we propose using graph attention network \cite{velivckovic2018graph} to incorporate the graph-structured data into learning the mapping from the parameters $\{ \bb{\theta}_i \}_{i \in \RRR}$ and $\{ \bb{\theta}_o \}_{o \in \OOO}$ to the optimal binary strategy for the edges, $\{ \bb{\delta}_{io} \}_{(i,o) \in \EEE_{RO}}$ and $\{ \bb{\delta}_{ij} \}_{(i,j) \in \EEE_R}$.

\begin{figure}[tb]
\centering
\includegraphics[width=0.5\textwidth, trim=160 160 250 170, clip]{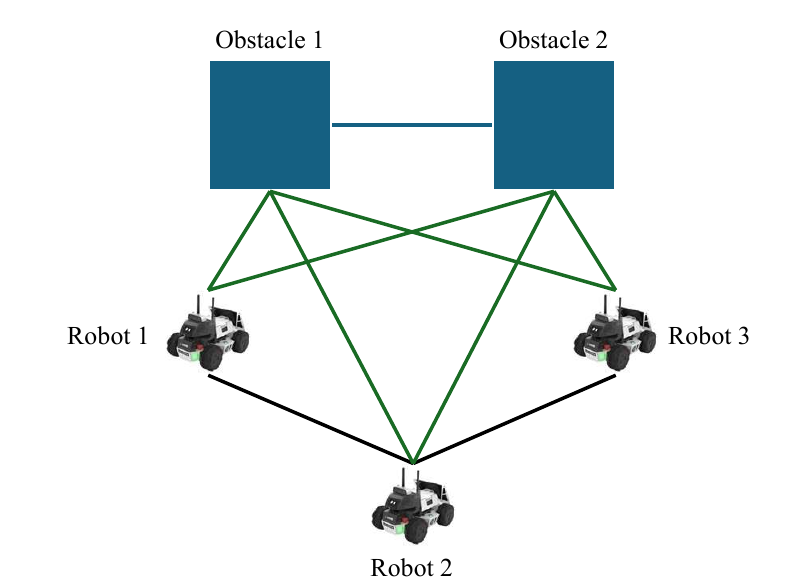}
\caption{Illustration of a heterogeneous graph for our problem. The robots and obstacles are represented as nodes, while the edges include robot-robot, robot-obstacle, and obstacle-obstacle connections.}
\label{fig:graph}
\vspace{-10pt}
\end{figure}

First, we model the heterogeneous multi-agent setting in our problem as a heterogeneous graph, where different types of agents and relations are represented by different types of nodes and edges.
Specifically, in our setting, the nodes include both robots and obstacles, while the edges represent relations between robot-robot and robot-obstacle pairs.
This definition of the heterogeneous graph is given as follows.
\begin{definition}
\label{def:graph}
(Graph) Let $\GGG = (\VVV, \EEE)$ be the graph to model the multi-robot system with obstacles, in which $\VVV = \RRR \cup \OOO$ and $\EEE = \EEE_R \cup \EEE_{RO} \cup \EEE_{OR} \cup \EEE_O$ be the node and edge sets, respectively, with $\EEE_{OR} = \{(o,i) \;|\; (i,o) \in \EEE_{RO}\}$ and $\EEE_O = \OOO \times \OOO$.
We include $\EEE_{OR}$ in the edge set to ensure that the graph is bidirectional.
The graph definition can be illustrated in Fig.~\ref{fig:graph}. 
\end{definition}

\subsection{Graph Attention Layers}

GAT network is a variant of the graph neural network (GNN), a deep learning model designed to process graph-structured data.
A fundamental component of a GAT network is \emph{graph attentional layer}.
Given input node features, \eg $(\bb{h}_1, \bb{h}_2, \dots, \bb{h}_N)$, and an adjacency matrix $\bb{A}_{\GGG}$, such that $\bb{A}_{mn} = 1$ if $(m, n) \in \EEE$, and $0$ otherwise,
a GAT layer computes a set of new node features, $(\bb{h}_1', \bb{h}_2', \dots, \bb{h}_N')$ where $N$ is the number of nodes.
In the initial step, a shared linear transformation, specified by a weight matrix $\bb{W}$ is applied to every node. 
This transforms the feature vectors into $\bb{g}_n = \bb{W} \bb{h}_n$. 
Then the importance of \node{m}'s features to \node{n} is described by attention coefficients $e_{mn}$ and its normalized value $\alpha_{mn}$  that are computed through an attentional mechanism $a$ and $\mathrm{softmax}$ function as follows
\begin{equation}
\begin{split}
e_{mn} &= a (\bb{g}_m, \bb{g}_n), \\
\alpha_{mn} &= \mathrm{softmax} (e_{mn}) = \frac{\exp(e_{mn})}{\sum_{p \in \NNN_n} \exp(e_{np})},
\end{split}
\end{equation}
Then the output for each node of a GAT layer is given by
\begin{equation}
\bb{h}_n' = \sigma \Big( \sum_{m \in \NNN_n} \alpha_{mn} \bb{W} \bb{h}_m \Big),
\end{equation}  
where $\sigma (.)$ is an activation function.
A GAT network is generally constructed by multiple GAT layers.
For more details on GAT networks, the readers are referred to \cite{velivckovic2018graph}.

\subsection{Heterogeneous Graph Attention Networks}

\begin{figure}[tb]
\centering
\includegraphics[width=0.48\textwidth, trim=5 110 50 100, clip]{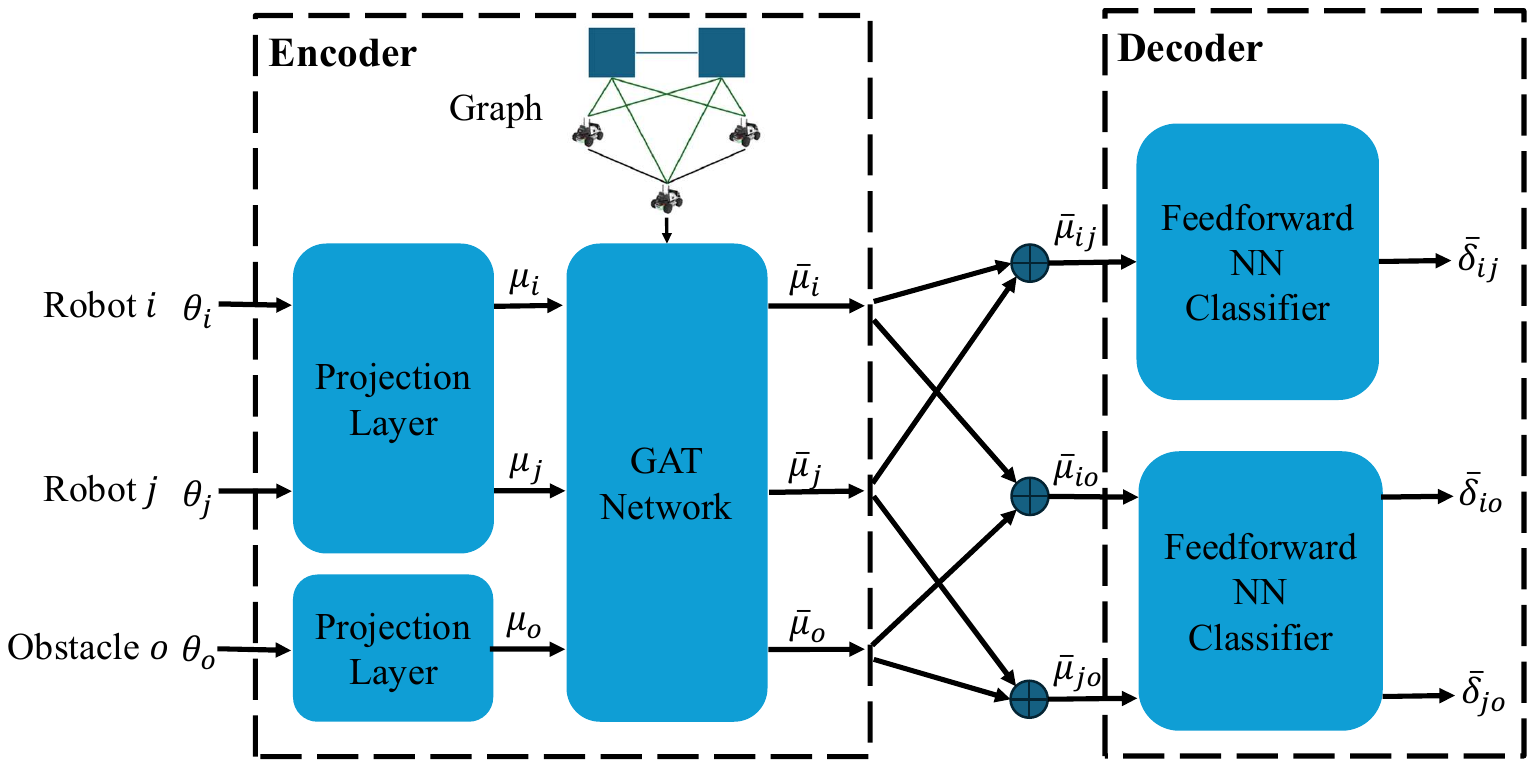}
\caption{Architecture of the neural network for learning optimal binaries.}
\label{fig:GATNN}
\vspace{-10pt}
\end{figure}

In this paper, we propose a NN architecture that can be illustrated in Fig.~\ref{fig:GATNN}.
This architecture consists of two modules: (1) a multi-layer heterogeneous GAT network (encoder) and (2) a feedforward NN classifier (decoder).
We utilize a multi-layer heterogeneous GAT network as an encoder to integrate graph-structured data, such as individual features of robots and obstacles and the connection links between them, and generate the node embeddings.
The heterogeneous GAT network \cite{wang2019heterogeneous} is an extension of the GAT network to handle the heterogeneity of graph data.
First, the features of nodes $\bbsym{\theta}_i$, $\forall\, i \in \VVV$ first pass through a projection layer (a linear transformation layer) to transform the features of different types of nodes into a uniform feature space.
The projection layer can be represented as follows
\begin{equation}
\begin{split}
\bbsym{\mu}_i &= \Phi_R\, \bbsym{\theta}_i, \forall\, i \in \RRR, \\
\bbsym{\mu}_o &= \Phi_O\, \bbsym{\theta}_o, \forall\, o \in \OOO,
\end{split}
\end{equation}
where $\Phi_R$ and $\Phi_O$ are matrices with appropriate dimensions.
Next, multiple GAT layers are used to aggregate the hidden features $\{\bbsym{\mu}_n\}$, $n \in \VVV$, while considering the relations of neighboring agents by the graph $\GGG$.  
Let $\bar{\bb{\mu}}_n$ be the output vector for each \node{n} $\in \VVV$ of the encoder, called node embeddings, then we have 
\begin{equation}
\{\bar{\bb{\mu}}_n\}_{n \in \VVV} = \Psi \Big( \{\bbsym{\mu}_n\}_{n \in \VVV}, \bb{A}_{\GGG} \Big),
\end{equation}
where $\Psi (\cdot)$ denotes the operator of the multi-layer GAT network.

The decoder, which takes the node embeddings as inputs, aims at predicting the optimal binary values $\{ \bb{\delta}_{io}^* \}_{(i,o) \in \EEE_{RO}}$ and $\{ \bb{\delta}_{ij}^* \}_{(i, j) \in \EEE_R}$, respectively.
For each edge in the heterogeneous graph, the input vector to the decoder is constituted by concatenating the embeddings of the two connecting nodes, \ie $\bar{\bb{\mu}}_{io}^\top = [\bar{\bb{\mu}}_i^\top, \bar{\bb{\mu}}_o^\top]$ for $(i,o) \in \EEE_{RO}$ and $\bar{\bb{\mu}}_{ij}^\top = [\bar{\bb{\mu}}_i^\top, \bar{\bb{\mu}}_j^\top]$ for $(i,j) \in \EEE_R$.
The concatenated embedding vector for each edge is passed through multi-layer feedforward NNs for classification.
Note that we use separate NNs for decoding the edges in $\EEE_{RO}$ and $\EEE_{R}$. 
We use standard feedforward NN classifiers with a ReLU activation function, as detailed in \cite{cauligi2020learning}, for the decoders, where we train the networks to predict the best values of the binary strategies.
The input-output relation of the decoder can be represented as follows:
\begin{equation}
\begin{split}
\bar{\bb{\delta}}_{io} &= \Omega_{RO} \big(\bar{\bb{\mu}}_{io} \big),\, \forall\, (i,o) \in \EEE_{RO}, \\
\bar{\bb{\delta}}_{ij} &= \Omega_{R} \big(\bar{\bb{\mu}}_{ij} \big),\, \forall\, (i,j) \in \EEE_{R},
\end{split}
\end{equation}
where $\bar{\bb{\delta}}_{io}$ and $\bar{\bb{\delta}}_{ij}$ are the output predictions of the decoder, while $\Omega_{RO} (\cdot)$ and $\Omega_{R} (\cdot)$ denote the decoder operators for edges in $\EEE_{RO}$ and $\EEE_{R}$, respectively.

\subsection{Data Generation and Training}

For offline data generation, we conducted large-scale simulations with numbers of robots ranging from $2$ to $5$, where the initial states of the robots and goal positions were randomly sampled.
We collect the data by solving the MICP using \texttt{GUROBIPY} solver \cite{gurobi}. 
If the problem is feasible, the problem parameters of the agents, the graph, and binary solutions are appended to the dataset $\DDD$.
Note that the MICP presented in Section~\ref{sec:prb} is not completely well-posed, in the sense that there may exist multiple values for the binary solution given a unique global solution of the continuous variable \cite{cauligi2021coco}. 
For instance, when considering the inter-robot collision avoidance constraint \eqref{eq:robot_ca}, if the optimal continuous solution satisfies $p^x_i(k) - p^x_j(k) \ge 2 d_{\min}$ and $p^y_i(k) - p^y_j(k) \ge 2 d_{\min}$ simultaneously, then $[b_{m,ij} (k)]_{m=1,\dots,4}$ can take any of the following values: $[0,1,1,1]$, $[1,1,0,1]$, or $[0,1,0,1]$.
This behavior can result in an ill-posed supervised learning problem, meaning that an input of the NN may correspond to multiple possible target outputs.
Therefore, given the solution obtained by \texttt{GUROBI} solver, we need to refine the binary solutions by the following rule:
if a constraint in \eqref{eq:robot_ca} or \eqref{eq:obs_ca} is satisfied regardless of whether the associated binary variable is $0$ or $1$, we set the binary variable to $0$.

We collected approximately $60,000$ data points from simulations and separated $90\%$ of the dataset for training and the remaining $10\%$ for validation.
We then constructed the NN with a linear projection layer of $64$ neurons, a two-layer GAT network with $64$ neurons per layer, followed by a ReLU feedforward NN with four layers and $256$ neurons per layer.  
To train the network, we minimize the cross-entropy between the ground-truth $(\bbsym{\delta}^*_{ij}, \bbsym{\delta}^*_{io})$ and the prediction $(\bar{\bbsym{\delta}}_{ij}, \bar{\bbsym{\delta}}_{io})$ over the training samples.
In terms of validation, the trained GAT network achieves $96\%$ and $93\%$ accuracy for edges in $\EEE_{RO}$ and $\EEE_R$, respectively.

\begin{remark}
In this paper, we assume that the obstacles are stationary. 
However, the framework can be easily extended to cases where obstacle features are dynamic and can be sampled.
\end{remark}

\begin{remark}
The inaccuracies in the NN's predictions may lead to infeasibility in the resulting convex program. 
Several approaches have been proposed in the literature to address this issue, such as selecting multiple candidates for the binary variables \cite{cauligi2020learning} or applying an iterative resolve method \cite{cauligi2022prism}. 
In this paper, we use soft constraints with a max penalty function (or equivalently, slack variables) to find the optimal solution with minimal constraint violation.
\end{remark}

\section{Distributed Convex Optimization with Proximal Alternating Direction Method of Multipliers (ADMM)}
\label{sec:opt}

\begin{algorithm}[b!]
\caption{Distributed Proximal ADMM Algorithm}
\label{algo:Alg}
\begin{algorithmic}[1]  
\Require $t_{\mathrm{max}}$, $\epsilon$, $\rho$, $\beta$, $\gamma$
\For {$t = 1,2,\dots,t_{\mathrm{max}}$}
\State \Robot{i} solves the $x$-minimization problem \eqref{eq:x_min} in parallel to obtain $\bbsym{y}_i^{(t+1)}$ and $\bbsym{z}_{\NNN_i}^{(t+1)}$
\State \Robot{i} transmits $\bbsym{y}_i^{(t+1)}$ to \agent{j}, $j \in \NNN_i$
\State \Robot{i} receives $\bbsym{y}_i^{(t+1)}$ from \agent{j}, $j \in \NNN_i$, and construct $\bbsym{y}_{\NNN_i}^{(t+1)}$
\State \Robot{i} update the dual variables using \eqref{eq:dual}
\If {$\norm{\bbsym{y}_i^{(t+1)} - \bbsym{y}_i^{(t)}} \le \epsilon$, $\forall\, i \in \RRR$}
\State \Return $\bbsym{y}_i^{(t+1)}$, $\forall\, i \in \RRR$
\EndIf
\EndFor
\State \Return $\bbsym{y}_i^{(t_{\max})}$, $\forall\, i \in \RRR$
\end{algorithmic}
\end{algorithm}

Once the binary decision variables are found by the NN, the remaining problem becomes convex, allowing us to utilize a distributed algorithm to solve it. 
By exploiting parallel computation, a distributed approach can reduce solving time in the online control phase, making it particularly effective when the number of agents is large.
We use the proximal alternating direction method of multipliers (ADMM) since it fully supports parallel computation \cite{deng2017parallel}. 
For ease of exposition, we first rewrite the optimization problem in the following form with a separable objective and coupling constraints:
\begin{subequations}
\label{eq:compact-prb}
\begin{align}
\underset{\bb{y}_i \in \YYY_i, \forall\, i \in \RRR}{\minimize} & \; \sum_{i \in \RRR} F_i (\bb{y}_i), \label{eq:compact-prb-obj} \\
\text{subject to} & \;\;  \bb{A}_{i} \bb{y}_i \le \bb{b}_{i}, \, i \in \RRR, \label{eq:compact-prb-loc} \\
& \;\; \bb{C}_{i} \bb{y}_i + \bb{C}_{\NNN_i} \bb{y}_{\NNN_i} \le \bb{d}_{i}, \label{eq:compact-prb-cpl}
\end{align}
\end{subequations}
where $\bb{y}_i$ is the local vector that collects all the continuous optimization variables of \agent{i}, and $\bb{y}_{\NNN_i}$ is the vector concatenating the variables of all the neighbors of \agent{i}.
Note that in \eqref{eq:compact-prb}, we consider the collision avoidance with obstacles as local constraints for each robot \eqref{eq:compact-prb-loc} and the inter-robot collision avoidance as the coupling constraints \eqref{eq:compact-prb-cpl}.
The problem \eqref{eq:compact-prb} can be rewriten equivalently by introducing an auxiliary variable $\bb{z}_{\NNN_i}$ as a local copy of $\bb{y}_{\NNN_i}$, leading to the following formulation:
\begin{equation}
\label{eq:admm-prob}
\begin{split}
\underset{\substack{\bb{y}_i \in \YYY, \bb{z}_{\NNN_i},\\ i \in \RRR}}{\text{minimize}} &\quad \sum_{i \in \RRR} F_i(\bb{y}_i) \\
\text{subject to} & \;\;  \bb{A}_{i} \bb{y}_i \le \bb{b}_{i}, \, i \in \RRR, \\
& \;\; \bb{C}_{i} \bb{y}_i + \bb{C}_{\NNN_i} \bb{z}_{\NNN_i} \le \bb{d}_{i}, \, i \in \RRR, \\
& \;\; \bb{y}_{\NNN_i} = \bb{z}_{\NNN_i},\; \forall\, i \in \RRR \text.
\end{split}
\end{equation} 
The augmented Lagrangian for the problem with equality constraints is formulated as follows
\begin{equation}
\begin{multlined}
\LLL \big( \{\bbsym{y}_i, \bbsym{z}_{\NNN_i}, \bbsym{\lambda}_i\}_{i\in \RRR} \big) 
=\! \sum_{i\in \RRR} F_i (\bbsym{y}_i) \\
+ \Big< \bbsym{\lambda}_i, \bbsym{z}_{\NNN_i} - \bbsym{y}_{\NNN_i} \Big> 
+ \frac{\rho}{2} \norm{\bbsym{z}_{\NNN_i} - \bbsym{y}_{\NNN_i}}^2 ,
\end{multlined}
\end{equation}
where $\bbsym{\lambda}_i$ is the vector of the dual variables (Lagrangian multipliers) associated with the equality constraint $\bb{y}_{\NNN_i} = \bb{z}_{\NNN_i}$, and $\rho > 0$ is a positive penalty weight.
The proximal ADMM algorithm \cite{deng2017parallel} consists of the following steps.
At each iteration $t$, each \agent{i} solves the following minimization problem in parallel:
\begin{equation}
\label{eq:x_min}
\begin{split}
\bbsym{y}_i^{(t+1)}, \bbsym{z}_{\NNN_i}^{(t+1)} = &\underset{\bb{y}_i \in \YYY, \bb{z}_{\NNN_i}}{\argmin} \; F_i(\bbsym{y}_i) + \frac{\beta}{2} \norm{\bbsym{z}_{\NNN_i} - \bbsym{z}_{\NNN_i}^{(t)}}^2  \\
& \hspace{4mm} + \frac{\rho}{2} \norm{\bbsym{z}_{\NNN_i} - \bbsym{y}_{\NNN_i}^{(t)} + \frac{\bbsym{\lambda}_i^{(t)}}{\rho}}^2, \\
\subjectto
& \quad \bbsym{A}_{i} \bbsym{y}_i \le \bbsym{b}_i, \\
& \quad \bb{C}_{i} \bb{y}_i + \bb{C}_{\NNN_i} \bb{z}_{\NNN_i} \le \bb{d}_{i},
\end{split}
\end{equation} 
where $\beta > 0$ is a positive weight for regularization.
Next, \agent{i} transmits $\bbsym{y}_i^{(t+1)}$ to \agent{j}, $j \in \NNN_i$.
After receiving all the information from the neighbors to construct $\bb{y}_{\NNN_i}^{(t+1)}$, \agent{i} updates the dual variables as follows
\begin{equation}
\label{eq:dual}
\bbsym{\lambda}_i^{(t+1)} = \bbsym{\lambda}_i^{(t)} + \gamma \rho\, \Big(\bbsym{z}_{\NNN_i}^{(t+1)} - \bbsym{y}_{\NNN_i}^{(t+1)}\Big),
\end{equation}
where $\gamma > 0$ is a damping coefficient. 
The iterations are repeated until convergence with a tolerance $\epsilon > 0$ or when a maximum number of iterations $t_{\max}$ is reached.
The algorithm can also be summarized in Algorithm~\ref{algo:Alg}.
Note that each computation step can be executed in parallel by the robots. 
In convex optimization, the proximal ADMM algorithm is guaranteed to converge under mild conditions on the parameter choice \cite{deng2017parallel}.

\section{Results and Discussions}

In this section, we validate the performance of our proposed framework with extensive simulations and experiments on a physical multi-robot testbed.

\subsection{Experiment Setup}

\begin{figure}[t]
\vspace{5pt}
\centering
\includegraphics[width=0.7\linewidth]{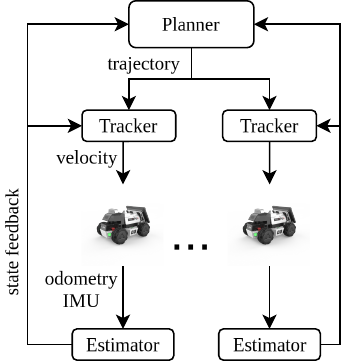}
\caption{The architecture of our experiment setup.}
\label{fig:architecture}
\vspace{-10pt}
\end{figure}

We validate the framework through experiments using the \texttt{LIMO} robotic testbed \cite{limo}.
The main program of our framework is implemented in Python, utilizing the PyTorch machine learning library for training and prediction of the NN model. 
The proximal ADMM algorithm is implemented in Julia, in which we created different threads for the robots to exploit parallel computation.
Integration between the proximal ADMM algorithm and the main program is achieved through the \texttt{PyCall} package.
We utilize the framework as a high-level trajectory planner to derive the optimal trajectories for the robots at every time step.
Since the \texttt{LIMO} robots are actuated with longitudinal and yaw velocity commands, \ie differentially driven robots, a lower-level tracking scheme, which is summarized next, is needed to compute these commands based on the planner's trajectory.
In all experiments, the control horizon length is chosen as $H=20$ and the trajectory planner is executed with a period of $\SI{200}{ms}$, while each tracker's period is $\SI{10}{ms}$.
The other parameters of the MICP are chosen as 
$v_{\max} = \SI{0.5}{m/s}$, $u_{\max} = \SI{0.5}{m/s^2}$, $d_{\min} = \SI{0.2}{m}$,
$\omega_u=1.0$, $\omega_p=1.0$, $\omega_{pt}=10.0$.
An extended Kalman filter \cite{moore2014generalized} is used for localization of each \texttt{LIMO} robot, which fuses odometry and IMU data to provide a sufficiently accurate state estimation.
The experiment setup can be also illustrated in Fig. \ref{fig:architecture}. 

To track the planner's trajectory, we implement the input-output linearization scheme proposed by Siciliano \etal \cite{siciliano2008}. %
Let $y_{i,1}$, $y_{i,2}$, $i \in \RRR$ be the outputs of the unicycle model, which are considered to be the Cartesian coordinates of a center point $B$ located along its sagittal axis, at a distance $b$ from the contact point of the wheel with the ground.
Those coordinates are computed as follows
\begin{equation}
\begin{split}
\label{eq:cartesian}
y_{i,1} & = p^x_i + b\,\cos\psi_i, \\
y_{i,2} & = p^y_i + b\,\sin\psi_i,
\end{split}            
\end{equation}
where $\psi_i$, $p^x_i$, $p^y_i$, are the robot's $i$ yaw angle and $x$, $y$ position with respect to a fixed global frame, respectively. 
The longitudinal and yaw velocity commands, $v_i^{\mathrm{cmd}},\omega_i^{\mathrm{cmd}}$, are then computed as
\begin{equation}
\begin{bmatrix}
v_i^{\mathrm{cmd}} \\
\omega_i^{\mathrm{cmd}}
\end{bmatrix}
=
\bb{T}^{-1}(\psi_i)
\begin{bmatrix}
u_{i,1} \\
u_{i,2}
\end{bmatrix},
\end{equation}
where $\bb{T}(\psi_i)$ is an invertible matrix and $u_{i,1}$, $u_{i,2}$ are the inputs of the following unicycle model:
\begin{align*}
\dot{y}_{i,1} & = u_{i,1}, \\
\dot{y}_{i,2} & = u_{i,2}, \\
\dot{\psi}_i & = \frac{u_{i,2}\,\cos\psi_i - u_{i,1}\,\sin\psi_i}{b}.
\end{align*}
Applying a linear control law, the control inputs can be given by
\begin{align*}
u_{i,1} & = \dot{y}_{i,1d} + k_{i,1}(y_{i,1d}-y_{i,1}), \\
u_{i,2} & = \dot{y}_{i,2d} + k_{i,2}(y_{i,2d}-y_{i,2}),
\end{align*}
where $k_{i,1} > 0$, and $k_{i,2} > 0$ are the gains of the linear feedback control law.
The linear control law guarantees exponential convergence to zero of the Cartesian tracking error. 
The desired variables $y_{i,1d}$, $y_{i,2d}$, $\dot{y}_{i,1d}$, $\dot{y}_{i,2d}$ are computed based on the solution obtained from the planner's optimal trajectory for the next time step $\{\bb{x}^*_{i}(t+1)\},\, i \in \RRR$.

\subsection{Experimental Validation}
\label{sec:exp}

\begin{figure*}[th]
\centering
\begin{minipage}{0.48\textwidth} %
    \centering
    \hspace{-6mm}
    \begin{subfigure}[b]{\textwidth}
        \includegraphics[width=1.1\textwidth]{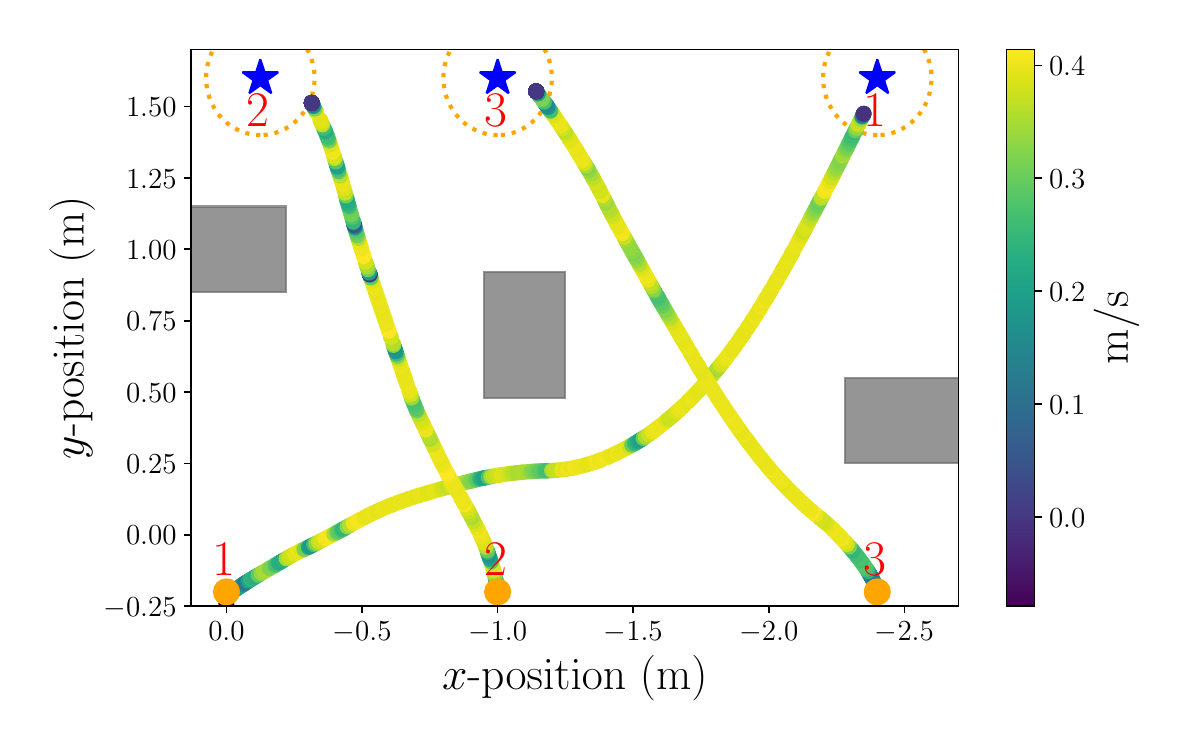}
        \caption{Position and velocity trajectories}
        \label{fig:trajectories}
    \end{subfigure}
\end{minipage}
~
\begin{minipage}{0.5\textwidth}
    \centering
    \begin{subfigure}[b]{0.48\textwidth} %
        \includegraphics[width=\textwidth]{figs/1.png}
        \caption{$t = 0 (s)$}
    \end{subfigure}
    ~
    \begin{subfigure}[b]{0.48\textwidth}
        \includegraphics[width=\textwidth]{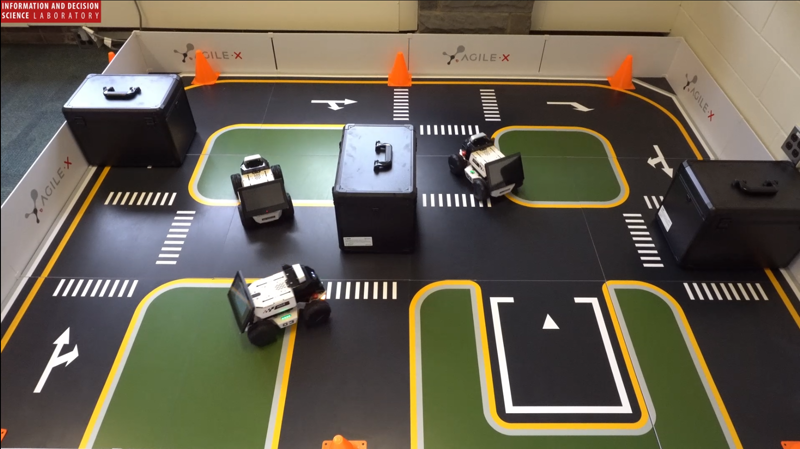}
        \caption{$t = 5 (s)$}
    \end{subfigure}

    \vspace{0.3cm} %

    \begin{subfigure}[b]{0.48\textwidth}
        \includegraphics[width=\textwidth]{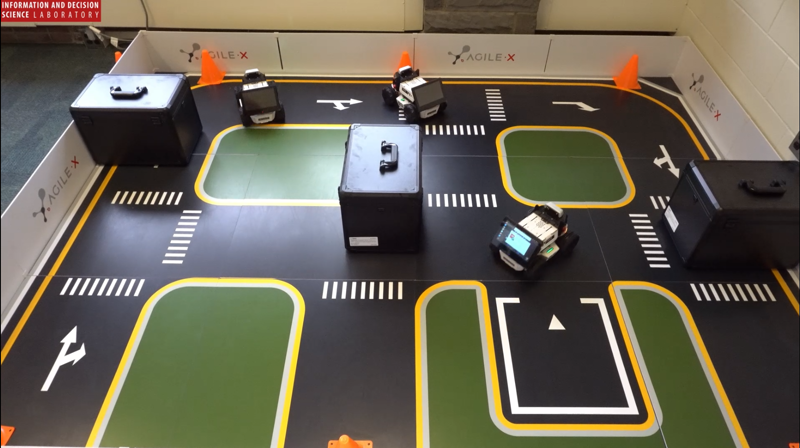}
        \caption{$t = 8 (s)$}
    \end{subfigure}
    ~
    \begin{subfigure}[b]{0.48\textwidth}
        \includegraphics[width=\textwidth]{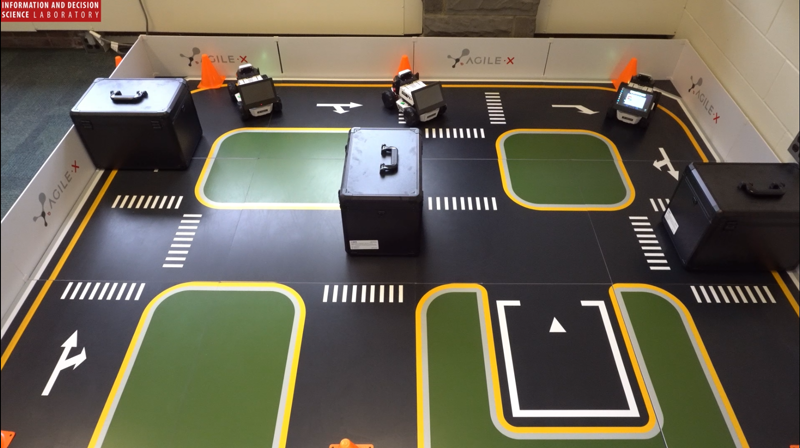}
        \caption{$t = 11 (s)$}
    \end{subfigure}
\end{minipage}

\caption{Trajectories of the robots and snapshots from an experiment with three robots and three obstacles.}
\label{fig:navigation}
\vspace{-10pt}
\end{figure*}

We conduct four experiments with three \texttt{LIMO} robots and three obstacles, each experiment corresponding to different goal assignments for the robots. 
Videos of the experiments can be found at \url{https://sites.google.com/cornell.edu/limo-micp}. 
Figure~\ref{fig:navigation} presents a specific experiment, illustrating the position and velocity trajectories of the robots, along with several snapshots captured during the experiment. 
Overall, the robots can reach their goals without any collisions.
The experiments verify that our proposed framework can be deployed in real time on a physical system.

\subsection{Simulation Studies}

To thoroughly evaluate the performance of our proposed framework, we conduct extensive randomized simulations with different numbers of robots to collect statistical results.
In particular, we analyze performance on the following metrics from the simulation data:
\begin{itemize}
\item \textbf{Success rate:} $\%$ of simulations in which all robots reach
their goals.
\item \textbf{Collision rate:} $\%$ of simulations that the collision avoidance constraints are violated.
\end{itemize}
We present the statistical results in Table~\ref{tab:metr}, which illustrate that as the number of robots increases, the success rate slightly decreases while the collision rate gradually rises. 
However, even with five robots in a highly constrained environment, the metrics remain relatively good.

\begin{figure}[tb]
\centering
\includegraphics[width=0.44\textwidth]{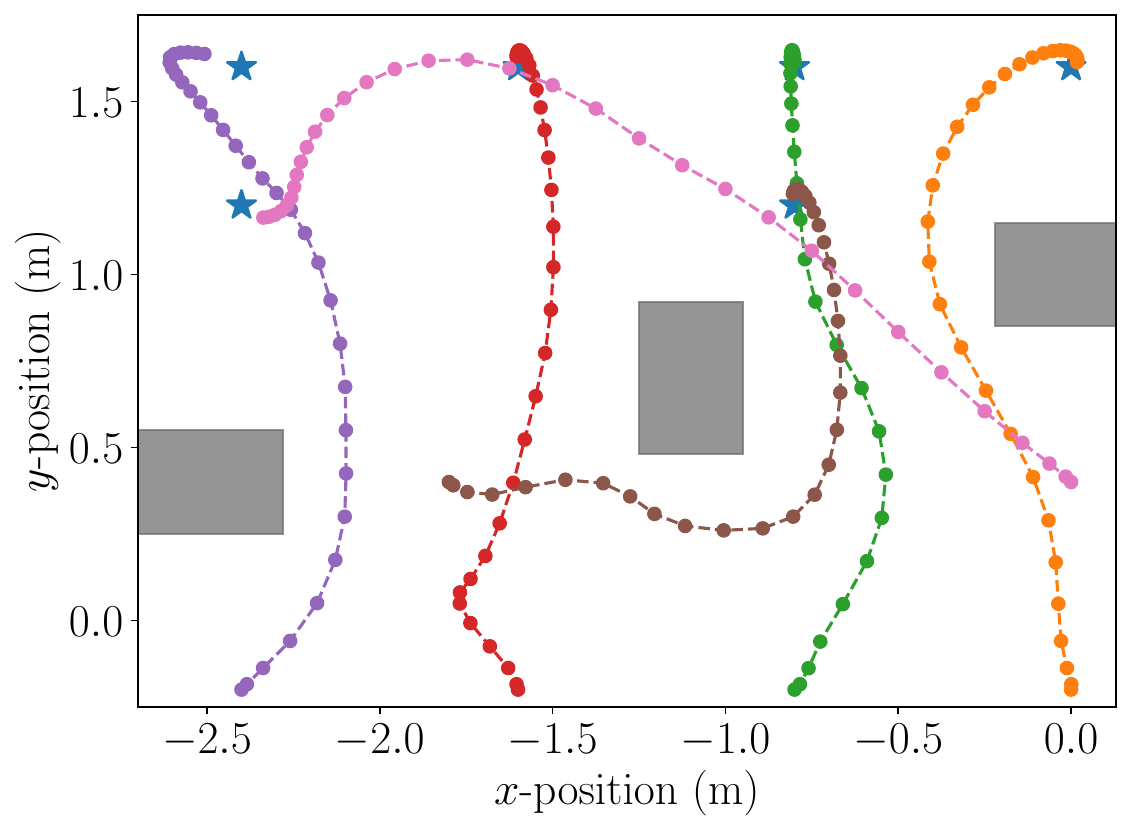}
\caption{Trajectories of the robots in the simulations with 6 robots. The obstacles are depicted as gray rectangles, while the robots' goals are indicated by stars.}
\label{fig:sim6}
\vspace{-10pt}
\end{figure}

Furthermore, an advantage of our framework is its ability to perform well when the number of robots in the test simulations differs from that in the training data. 
Specifically, while we train the model using a dataset generated from simulations with $2$ to $5$ robots, we additionally evaluate it in a simulation with $6$ robots.
The trajectories of the robots in this simulation, shown in Fig.~\ref{fig:sim6}, demonstrate that the robots can finish the navigation task safely.

\begin{table}
\centering
\caption{Statistical results for simulations with different number of robots.}
\label{tab:metr}
\begin{tabular}{ K{2.5cm} | K{0.8cm} | K{0.8cm} | K{0.8cm} | K{0.8cm} }
\toprule[1.2pt]%
Numbers of robots & 2 & 3 & 4 & 5 \\ 
\midrule[0.6pt]
Success rate (\%) 
& $100$ & $99.8$ 
& $99.5$ & $99.2$ \\
Collision rate (\%) 
& $0.0$ & $0.3$ 
& $0.8$ & $2.1$ \\
\bottomrule[1.2pt]
\end{tabular}
\end{table}

In terms of computation time, the distributed optimization algorithm has advantages over a centralized approach only when the number of robots is high. 
For example, in the simulation with six robots, our proposed framework took $0.15$ seconds, while the centralized \texttt{GUROBIPY} solver took $0.27$ seconds on average. 
This computation time includes both the time for NN prediction and for solving convex programs.
In contrast, in a simulation/experiment with three robots, our algorithm required approximately the same computation time as \texttt{GUROBIPY}. 
Therefore, we believe that the proposed framework can be beneficial in reducing computation time when applied to more complex and large-scale multi-agent systems, such as traffic signal control for connected automated vehicles in mixed traffic \cite{le2024distributed}.

\section{Conclusions}
\label{sec:conc}

In this paper, we presented a comprehensive framework for solving a MICP arising from real-time multi-robot navigation. This framework leverages a heterogeneous GAT and employs proximal ADMM-based distributed optimization.
Extensive experimental validation and simulation results demonstrate that the proposed framework effectively enables robot navigation while maintaining reasonable computational time.
While our framework has been specifically applied to multi-robot navigation, it possesses the potential to be extended to more complex and large-scale multi-agent systems. Future research endeavors will explore this extension.

\balance

\bibliographystyle{IEEEtran}
\bibliography{IEEEabrv,refs,refs_IDS}

\end{document}

%% file: defs.tex
\newcommand{\node}[1]{node\textendash\ensuremath{#1}\xspace}
\newcommand{\agent}[1]{agent\textendash\ensuremath{#1}\xspace}

\newcommand{\robot}[1]{robot\textendash\ensuremath{#1}\xspace}
\newcommand{\Robot}[1]{Robot\textendash\ensuremath{#1}\xspace}
\newcommand{\obstacle}[1]{obstacle\textendash\ensuremath{#1}\xspace}

\newcommand{\bbsym}[1]{\ensuremath{\boldsymbol{#1}}}

%% file: main.bbl
\begin{thebibliography}{10}
\providecommand{\url}[1]{#1}
\csname url@rmstyle\endcsname
\providecommand{\newblock}{\relax}
\providecommand{\bibinfo}[2]{#2}
\providecommand\BIBentrySTDinterwordspacing{\spaceskip=0pt\relax}
\providecommand\BIBentryALTinterwordstretchfactor{4}
\providecommand\BIBentryALTinterwordspacing{\spaceskip=\fontdimen2\font plus
\BIBentryALTinterwordstretchfactor\fontdimen3\font minus
  \fontdimen4\font\relax}
\providecommand\BIBforeignlanguage[2]{{%
\expandafter\ifx\csname l@#1\endcsname\relax
\typeout{** WARNING: IEEEtran.bst: No hyphenation pattern has been}%
\typeout{** loaded for the language `#1'. Using the pattern for}%
\typeout{** the default language instead.}%
\else
\language=\csname l@#1\endcsname
\fi
#2}}

\bibitem{quirynen2024real}
R.~Quirynen, S.~Safaoui, and S.~Di~Cairano, ``Real-time mixed-integer quadratic
  programming for vehicle decision-making and motion planning,'' \emph{IEEE
  Transactions on Control Systems Technology}, 2024.

\bibitem{le2024distributed}
V.-A. Le and A.~A. Malikopoulos, ``{Distributed Optimization for Traffic Light
  Control and Connected Automated Vehicle Coordination in Mixed-Traffic
  Intersections},'' \emph{IEEE Control Systems Letters}, vol.~8, pp.
  2721--2726, 2024.

\bibitem{bang2021AEMoD}
H.~Bang and A.~A. Malikopoulos, ``Congestion-aware routing, rebalancing, and
  charging scheduling for electric autonomous mobility-on-demand system,'' in
  \emph{Proceedings of 2022 American Control Conference (ACC)}, 2022, pp.
  3152--3157.

\bibitem{camisa2022multi}
A.~Camisa, A.~Testa, and G.~Notarstefano, ``Multi-robot pickup and delivery via
  distributed resource allocation,'' \emph{IEEE Transactions on Robotics},
  vol.~39, no.~2, pp. 1106--1118, 2022.

\bibitem{le2023multirobot}
V.-A. Le, B.~Chalaki, V.~Tadiparthi, H.~N. Mahjoub, J.~D'sa, E.~Moradi-Pari,
  and A.~A. Malikopoulos, ``{Multi-Robot Cooperative Navigation in Crowds: A
  Game-Theoretic Learning-Based Model Predictive Control Approach},'' in
  \emph{2024 IEEE International Conference on Robotics and Automation
  (ICRA)}.\hskip 1em plus 0.5em minus 0.4em\relax IEEE, 2024, pp. 4834--4840.

\bibitem{salvado2018motion}
J.~Salvado, R.~Krug, M.~Mansouri, and F.~Pecora, ``Motion planning and goal
  assignment for robot fleets using trajectory optimization,'' in \emph{2018
  IEEE/RSJ International Conference on Intelligent Robots and Systems
  (IROS)}.\hskip 1em plus 0.5em minus 0.4em\relax IEEE, 2018, pp. 7939--7946.

\bibitem{bestehorn2019switching}
F.~Bestehorn, C.~Hansknecht, C.~Kirches, and P.~Manns, ``A switching cost aware
  rounding method for relaxations of mixed-integer optimal control problems,''
  in \emph{2019 IEEE 58th Conference on Decision and Control (CDC)}.\hskip 1em
  plus 0.5em minus 0.4em\relax IEEE, 2019, pp. 7134--7139.

\bibitem{zhang2021feasibility}
C.~Zhang, Z.~Dong, and L.~Yang, ``A feasibility pump based solution algorithm
  for two-stage robust optimization with integer recourses of energy storage
  systems,'' \emph{IEEE Transactions on Sustainable Energy}, vol.~12, no.~3,
  pp. 1834--1837, 2021.

\bibitem{stellato2016high}
B.~Stellato, T.~Geyer, and P.~J. Goulart, ``High-speed finite control set model
  predictive control for power electronics,'' \emph{IEEE Transactions on power
  electronics}, vol.~32, no.~5, pp. 4007--4020, 2016.

\bibitem{cauligi2020learning}
A.~Cauligi, P.~Culbertson, B.~Stellato, D.~Bertsimas, M.~Schwager, and
  M.~Pavone, ``Learning mixed-integer convex optimization strategies for robot
  planning and control,'' in \emph{2020 59th IEEE conference on decision and
  control (CDC)}.\hskip 1em plus 0.5em minus 0.4em\relax IEEE, 2020, pp.
  1698--1705.

\bibitem{cauligi2021coco}
A.~Cauligi, P.~Culbertson, E.~Schmerling, M.~Schwager, B.~Stellato, and
  M.~Pavone, ``Coco: Online mixed-integer control via supervised learning,''
  \emph{IEEE Robotics and Automation Letters}, vol.~7, no.~2, pp. 1447--1454,
  2021.

\bibitem{bertsimas2022online}
D.~Bertsimas and B.~Stellato, ``Online mixed-integer optimization in
  milliseconds,'' \emph{INFORMS Journal on Computing}, vol.~34, no.~4, pp.
  2229--2248, 2022.

\bibitem{srinivasan2021fast}
M.~Srinivasan, A.~Chakrabarty, R.~Quirynen, N.~Yoshikawa, T.~Mariyama, and
  S.~Di~Cairano, ``Fast multi-robot motion planning via imitation learning of
  mixed-integer programs,'' \emph{IFAC-PapersOnLine}, vol.~54, no.~20, pp.
  598--604, 2021.

\bibitem{cauligi2022prism}
A.~Cauligi, A.~Chakrabarty, S.~Di~Cairano, and R.~Quirynen, ``Prism: Recurrent
  neural networks and presolve methods for fast mixed-integer optimal
  control,'' in \emph{Learning for Dynamics and Control Conference}.\hskip 1em
  plus 0.5em minus 0.4em\relax PMLR, 2022, pp. 34--46.

\bibitem{bertsimas2023prescriptive}
D.~Bertsimas and C.~W. Kim, ``A prescriptive machine learning approach to
  mixed-integer convex optimization,'' \emph{INFORMS Journal on Computing},
  vol.~35, no.~6, pp. 1225--1241, 2023.

\bibitem{russo2023learning}
L.~Russo, S.~H. Nair, L.~Glielmo, and F.~Borrelli, ``Learning for online
  mixed-integer model predictive control with parametric optimality
  certificates,'' \emph{IEEE Control Systems Letters}, vol.~7, pp. 2215--2220,
  2023.

\bibitem{tang2024learning}
B.~Tang, E.~B. Khalil, and J.~Drgo{\v{n}}a, ``Learning to optimize for
  mixed-integer non-linear programming,'' \emph{arXiv preprint
  arXiv:2410.11061}, 2024.

\bibitem{deng2017parallel}
W.~Deng, M.-J. Lai, Z.~Peng, and W.~Yin, ``Parallel multi-block admm with o
  (1/k) convergence,'' \emph{Journal of Scientific Computing}, vol.~71, pp.
  712--736, 2017.

\bibitem{Malikopoulos2021}
A.~A. Malikopoulos, ``On team decision problems with nonclassical information
  structures,'' \emph{IEEE Transactions on Automatic Control}, vol.~68, no.~7,
  pp. 3915--3930, 2023.

\bibitem{alrifaee2014centralized}
B.~Alrifaee, M.~G. Mamaghani, and D.~Abel, ``Centralized non-convex model
  predictive control for cooperative collision avoidance of networked
  vehicles,'' in \emph{2014 IEEE international symposium on intelligent control
  (ISIC)}.\hskip 1em plus 0.5em minus 0.4em\relax IEEE, 2014, pp. 1583--1588.

\bibitem{velivckovic2018graph}
P.~Veli{\v{c}}kovi{\'c}, G.~Cucurull, A.~Casanova, A.~Romero, P.~Li{\`o}, and
  Y.~Bengio, ``Graph attention networks,'' in \emph{International Conference on
  Learning Representations}, 2018.

\bibitem{wang2019heterogeneous}
X.~Wang, H.~Ji, C.~Shi, B.~Wang, Y.~Ye, P.~Cui, and P.~S. Yu, ``Heterogeneous
  graph attention network,'' in \emph{The world wide web conference}, 2019, pp.
  2022--2032.

\bibitem{gurobi}
\BIBentryALTinterwordspacing
{Gurobi Optimization, LLC}, ``Gurobi optimizer reference manual,'' 2021.
  [Online]. Available: \url{http://www.gurobi.com}
\BIBentrySTDinterwordspacing

\bibitem{limo}
\BIBentryALTinterwordspacing
{Agilex Robotics}. [Online]. Available:
  \url{https://global.agilex.ai/products/limo-pro}
\BIBentrySTDinterwordspacing

\bibitem{moore2014generalized}
T.~Moore and D.~Stouch, ``A generalized extended kalman filter implementation
  for the robot operating system,'' in \emph{Proceedings of the 13th
  International Conference on Intelligent Autonomous Systems (IAS-13)}.\hskip
  1em plus 0.5em minus 0.4em\relax Springer, 2014, pp. 335--348.

\bibitem{siciliano2008}
B.~Siciliano, L.~Sciavicco, L.~Villani, and G.~Oriolo, \emph{Robotics:
  modelling, planning and control}.\hskip 1em plus 0.5em minus 0.4em\relax
  Springer, 2008.

\end{thebibliography}
